\documentclass[3p,twocolumn]{elsarticle}
\usepackage{graphicx}
\usepackage{bm}
\usepackage{amsmath,amssymb}
\begin{document}
\title{Universality classes for Coulomb frustrated phase separation }

\author[lei]{C. Ortix\corref{cor1}}
\ead{ortix@lorentz.leidenuniv.nl}

\author[sap,isc]{J. Lorenzana}
\ead{jose.lorenzana@roma1.infn.it}

\author[sap]{C. Di Castro}
\ead{carlo.dicastro@roma1.infn.it}

\address[lei]{Insitute-Lorentz for Theoretical Physics, Universiteit Leiden, P.O. Box 9506, 2300 RA Leiden, The Netherlands}

\address[sap]{SMC-INFM-CNR and Dipartimento di Fisica, Universit\`a di Roma 
``La Sapienza'', P.  Aldo Moro 2, 00185 Roma, Italy}

\address[isc]{ISC-CNR, Via dei Taurini 19, 00185 Roma, Italy} 

\cortext[cor1]{Corresponding author. Tel.: +31 715275540; Fax: +31 715275511}

\begin{abstract}
We identify two ``universality'' classes in the Coulomb frustrated phase separation phenomenon. They correspond to two different kind of electronic compressibility anomalies often encountered in strongly correlated electronic systems. We discuss differences and similarities of their corresponding phase diagrams in two- and three-dimensional systems. 
\end{abstract}

\maketitle

\begin{section}{Introduction}

A large variety of systems with phase separation (PS) tendencies
subject to long-range forces self-organize in domain patterns
\cite{seu95,sci04,kit46,oht86}.  
Recently, advances in local probe techniques have revealed mixed
states in materials like cuprates and manganites \cite{lan02,bec02}
rekindling the study of this phenomenon in strongly correlated
electronic systems. Indeed it has become clear that strong electron
correlations generally produce a tendency towards PS
\cite{cas92,cas95b,low94,eme90c,nag98}   which, however, is frustrated by the long-range part of the Coulomb interaction (LRC). This leads to the formation of
inhomogeneities with a typical size determined by the 
competition among long range forces and surface energy effects.

Since domains have often mesoscopic scales of several lattice
constant, one can perform a general analysis of the frustrated phase
separation (FPS) mechanism which neglects the microscopic details of
each specific system while capturing its general properties. Tendency towards
PS is then recognized by the presence of anomalies in the electronic
contribution to the free energy of the system \cite{ort07b}. The
anomalies found in a large variety of 
strongly correlated electronic models can be classified in two kinds 
corresponding to a short-range negative compressibility
density region \cite{nag67,kag99,kug05,don95} or a Dirac-delta-like
negative divergence of the compressibility due to the crossing of the
free energies of two distinct phases \cite{oka00,bri99b}. Both
situations can be captured by expanding the short range part of the  
electronic free energy
density $f_e$ of the system around a reference density $n_c$ as
$f_e=\alpha |n-n_c|^{\gamma}$. Here $\alpha<0$ encodes the tendency towards
PS where  $\gamma=1$ corresponds to a compressibility
divergence and $\gamma=2$ to a negative compressibility region.  Higher
order terms are essential to analyze FPS from the limit of
strong frustration down to the limit of zero frustration. As a minimal
model we take a contribution to the free energy of the form 
$\beta (n-n_c)^{2 \gamma}$ that for $\beta>0$ provide
a symmetric double-well form of the short range free energy.  

When LRC effects can be considered as a weak perturbation upon the
ordinary PS mechanism, one can achieve a universal picture of the FPS
 \cite{ort07b}. On the contrary, in the strongly frustrated
regime, the two short-range compressibility anomalies give rise to two
different behaviors. The aim of this work is to review the foremost
features of their corresponding phase diagrams for two- ($D=2$) and
three-dimensional ($D=3$) systems embedded in the three-dimensional
long-range Coulomb interaction.  
\end{section}

\begin{section}{Universal behavior: the weak frustration regime}

In the $\gamma=2$ case, FPS can be analyzed by means of a paradigmatic $\phi^4$ model augmented with the long-range Coulomb interaction. Pattern formation within this model  (or closely related variants)  has been
considered in a variety of systems \cite{ort07bis} including mixtures of block copolymers \cite{oht86}, charged colloids in polymeric solutions \cite{tar06} and electronic systems \cite{low94,sch00}. 
The corresponding Hamiltonian reads:  
\begin{eqnarray}
  {\cal H}&=&\int d{\bf x} \left[\phi^{2}({\bf x})-1\right]^2+|\nabla \phi({\bf x})|^2+\frac{Q^2}{2} \nonumber \\  & \times &  \int d{\bf x} \int d{\bf x^{\prime}}\dfrac{\left[\phi({\bf x})-\overline{\phi}\right] \left[\phi({\bf x^{\prime}})-\overline{\phi}\right]}{|{\bf x-x^{\prime}}|} 
\label{eq:modelgamma2}
\end{eqnarray}
where the classical scalar field $\phi$ represents the local charge density with $\overline{\phi}$ its average density. A rigid background ensures charge neutrality.  
The model Eq.~(\ref{eq:modelgamma2}) can be reached by measuring the electronic free energy density $f_e$ in unit of the barrier height $\alpha^2/(4 \beta)$, densities in units such that the double-well minima are at $\phi=\pm 1$, 
and distances in unit of the bare correlation length $\xi$. This leads
to a renormalized Coulomb coupling $Q^2=2 e^2 \xi^{D-1}/(\varepsilon_0
|\alpha|)$ with $e$ being the electronic charge and $\varepsilon_0$ a dielectric constant due to external degrees of freedom. 

Frustrated phase separation for $\gamma=1$
is more easily described by adding
an auxiliary field $s$, equivalent to an Hubbard-Stratonovich variable, linearly coupled to the charge. It can be taken as a soft \cite{jam05} or a conventional
Ising spin ($s=\pm 1$) with the sign distinguishing the two competing phases. In the remainder we will refer to the latter case which is more
straightforward to analyze. Domain
walls of the Ising order parameter are sharp by construction with a
surface tension $\sigma \propto 2 J$  where $J$ indicates the Ising coupling.
Then, in the continuum limit, one obtains the following Hamiltonian:
\begin{eqnarray}
{\cal H}&=& \Sigma+\int d {\bf x} [\phi({\bf x})-s({\bf x})]^{2}+  \frac{Q^2}2   \nonumber \\
 & \times  &\int d {\bf x}\int d {\bf x^{\prime}}
 \dfrac{\left[\phi\left({\bf x}\right)- \overline{\phi}\right]
\left[\phi \left({\bf x^{\prime}}\right)- \overline{\phi}\right]}{\left|{\bf x-x^{\prime}}\right|} 
\label{eq:modelgamma1}
\end{eqnarray}
As before, we measure energy densities in units of $
\alpha^2 /(4 \beta)$, the minima of the double well are at $\phi=\pm 1$ and lengths are measured in units of $\xi \equiv 4\sigma \beta/\alpha^2$ which represents the analogue of the bare correlation length in the present model. More precisely it indicates the size that
inhomogeneities should have for the total interface energy to 
be of the same order as the phase separation energy density gain 
$\alpha^2 /(4 \beta)$. 
As for the model Eq.~(\ref{eq:modelgamma2}), the phase diagram is spanned by the dimensionless 
average density $\overline{\phi}$ and the renormalized Coulomb coupling $Q^2= e^2 \xi^{D-1}/(\varepsilon_0 \beta)$.

In the absence of the Coulomb interaction, both models are subject to
ordinary PS in the density range $|\overline{\phi}|<1$ as ruled by the Maxwell construction (MC).
Hence, the mixed state consists of macroscopic domains with local densities $\phi=\pm 1$. 
For $Q \neq 0$, PS is undermined as a thermodynamic phenomenon since the LRC energy cost grows faster than the volume in the
thermodynamic limit. Thus charged domains at a mesoscopic scale appear. 
Their typical size $l_d$ is determined by the competition between the
LRC cost $\sim Q^2 l_d^{D-1}$ and the surface energy density $\sim
l_d^{-1}$. These terms are optimized whenever the inhomogeneities get
a typical size $ l_d^D \sim 1/Q^2$.

Another important length scale is the screening length of the Coulomb
interaction that controls the relaxation of the electronic charge inside the domains. Both the models Eqs.~(\ref{eq:modelgamma2}),~(\ref{eq:modelgamma1}), have a characteristic screening length defined by $l_{s}^{D-1} \sim 1/Q^{2}.$
It thus follows that in the weak frustration regime $Q<<1$, one
obtains the hierarchy of length scales (in units of $\xi$): $l_s>>l_d>>1$ that gives ground for a unified treatment of the FPS mechanism. 

The strong separation between the typical size of the domains and the typical interface width $\sim 1$  allows to consider the smooth interface of the model Eq.~(\ref{eq:modelgamma2}) as sharp with a surface tension defined as  the excess energy of an isolated
interface \cite{mur02}. Moreover, since the effect of LRC represents a small perturbation upon ordinary PS, inhomogeneities will appear with local densities near $\phi=\pm 1$. Then, FPS can be analyzed by expanding quadratically the double-well potential of Eq.~(\ref{eq:modelgamma2}) around its minima thus leading to a practical equivalence among the two FPS models.

A good approximation in the weakly frustrated regime lies in assuming a uniform density approximation (UDA) in which 
 the local charge density 
is assumed constant\cite{lor01I,lor01II,lor02,ort06,ort07}. The resulting phase diagram in this approximation is shown with the thick lines at the bottom of Fig.~\ref{eq:figgamma2} for the $\gamma=2$ case in 3D systems. 
Comparison with exact results shows 
that the UDA gives a  very accurate description. 
From Fig.~\ref{eq:figgamma2} one sees that droplet-like domains are
 the stable morphologies on entering in the inhomogeneous region. As 
$\overline\phi$ approaches the origin at fixed $Q$ a 
topological transition to rod-like structures and subsequently to
 layered structures occurs. A similar behavior is also expected in 2D
 systems. 

Our computations at weak coupling are variational so we can not exclude more
complicated phases including elongation of domains and ``fingering''  
as in classical systems \cite{seu95}.  Naturally the ordered phases will
be very sensitive to quenched disorder.  Also in the absence of 
quenched disorder the ground state may be hard to reach on a cooling
experiment leading to a glass state.\cite{sch00}

\begin{figure}
\includegraphics[width=7cm]{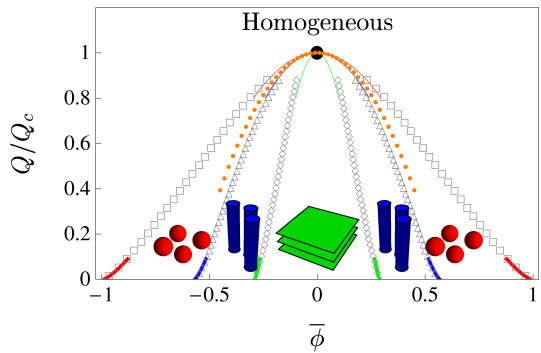}
\caption{The $\gamma=2$ phase diagram in three-dimensional systems. The small dots indicates the Gaussian instability line $Q_g$. The thin (thick) lines represent first-order transitions in the strong (weak) coupling approximation. In the two limits they overlap with the corresponding numerically determined transition lines from the homogeneous phase to droplet inhomogeneities ($\square$), from droplets to rods ($\triangle$) and from rods to layers ($\diamondsuit$). Finally the black circle indicates the CP.}
\label{eq:figgamma2}
\end{figure}

\end{section}

\begin{section}{Universality classes: the strong frustration regime}

By increasing the renormalized Coulomb coupling in Eqs.~(\ref{eq:modelgamma2}),~(\ref{eq:modelgamma1}), inhomogeneous states  with local densities close to the reference density  emerge. In this
case the behavior of the two FPS models is radically different  leading to
two ``universality'' classes.
 
First, we consider the $\gamma=2$ model Eq.~(\ref{eq:modelgamma2}). We restrict to 3D systems but similar ideas applies also to the 2D case. By computing the static response to an external field in momentum space, we get the charge susceptibility at ${\bf k} \neq 0$ as:
$$\chi({\bf k})=\left[{\bf k}^2+\dfrac{2 \pi Q^{2}}{{\bf k}^2}-2+6 \overline{\phi}^{2}\right]^{-1}.$$
 The charge susceptibility has a maximum at the characteristic finite wavevector $k_0=\left[2 \pi Q^{2}\right]^{1/4}$ and diverges approaching the Gaussian instability line [dotted line in Fig.~\ref{eq:figgamma2}]  $Q_g=Q_c(1-3 \overline{\phi}^{2})$ where $Q_c=1/\sqrt{2 \pi}$. This indicates an instability towards a sinusoidal charge density wave (SCDW) with direction chosen by spontaneous symmetry breaking. 
At small $Q$ the Gaussian transition line  predicts inhomogeneities
 within the spinodal region $\left|\overline{\phi}\right|<
 1/\sqrt{3}$. This contrast with ordinary PS at $Q=0$ which implies a
 mixed state in the global  
density region $\left|\overline{\phi}\right|<1$. 
 The situation has been recently
 clarified in Ref. \cite{ort07bis} where it has been shown that
 inclusion of non-Gaussian terms results in a first-order phase
 transition preempting the second-order Gaussian instability line
 except for the critical point (CP)
 $\left(\overline{\phi},Q\right)=\left(0, Q_c \right)$. The mixed
 region smoothly connects with the macroscopically phase separated
 state.  

Away but close to the CP, the transition is weakly first-order with more complicated morphologies. 
Approaching the first-order line from above, inhomogeneities are predicted to form a BCC lattice with subsequent topological transitions to a planar hexagonal lattice of rods and layered structures [see Fig.~\ref{eq:figgamma2} ]. One then finds 
the same topology as in the weak coupling regime, but now inhomogeneities have smooth interfaces in between. 
They continuously evolve into sharply defined droplets, rods and
layers (disks and stripes in 2D)  as $Q \rightarrow 0$ with a
proliferation of higher order harmonics.   

\begin{figure}
\includegraphics[width=7cm]{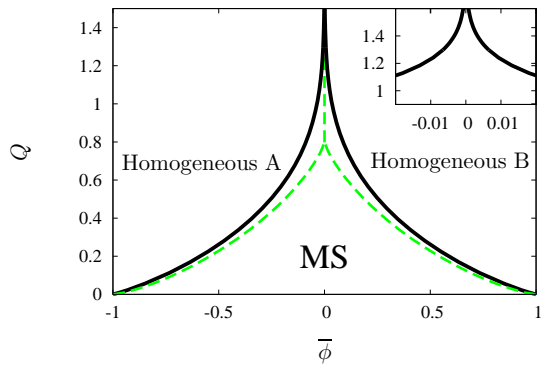}
\caption{The $\gamma=1$ phase diagram in D=2 (full line) and $D=3$ (dashed line) for the {\it smectic} solution. The central region corresponds to the mixed state (MS). The inset shows an enlargement at strong frustration near the logarithmic singularity of the two-dimensional phase diagram.}
\label{eq:figgamma1}
\end{figure}

Next, we analyze the $\gamma=1$ universality class. Fig.~\ref{eq:figgamma1} shows the phase diagrams for two- and three-dimensional systems 
respectively for striped and layered structures which describe  a {\it
  smectic} electronic liquid phase that possesses orientational order
and breaks the translational symmetry only in one direction. As for
the previous model other morphologies will be competitive away from 
$\overline\phi=0$. Indeed at weak coupling, as mentioned above, the two
phases diagrams are identical. Here, however, we are interested in the
behavior approaching $\overline\phi=0$ where the considered morphology
is the most stable one.

In 3D systems, one finds a maximum frustration degree above which only uniform phases are allowed. This is antithetical to the behavior of 2D systems where the transition lines diverge
logarithmically at $\overline\phi=0$ \cite{jam05,ort07} and thus the
system always breaks into domains no matter how strong the frustrating
effects are. 

The difference between 2D and 3D results can be traced
back to the different role of screening for different
dimensionalities. 
In three dimensions, the charge density decays exponentially from the domain interfaces on the scale of the Thomas-Fermi screening length whereas in 2D systems it decays as a power law. Phase
separation energy gain stems from the region where the electronic
density is significantly different from its average value
\cite{lor01I}. In two-dimensions this is fulfilled everywhere in the
domains, even far from the boundaries. This allows for domains with
any 
typical size, even exponentially larger than the screening length. On the
contrary, in three-dimensions the systems gain PS energy in a region
of  width $l_s$ around the interface. Regions far from the boundary
produce an exponentially small energy gain and thus the system adjust
itself to eliminate them.  As a consequence a maximum
size rule is generally valid that says that inhomogeneities cannot have all linear
dimensions much larger than the three-dimensional screening length
$l_s^{3D}$. This allows for arbitrary large inhomogeneities in 2D
systems since one of the dimension is already smaller than the
$l_s^{3D}$.

\end{section}

\begin{section}{Conclusions}

In this work we reviewed the main features of frustrated phase separation in charged systems considering two kind of short-range compressibility anomalies often encountered in strongly correlated electronic systems.  
The effect of long-range forces can be measured by introducing a
dimensionless renormalized Coulomb coupling $Q$ which is a measure of the amount of frustration.  Frustration tends to reduce the range of
density where a mixed state appears hence stabilizing the homogeneous phase at densities where ordinary PS would occur. 
 This situation is in accord with  thermodynamic 
measurements \cite{eis92} of the uniform two-dimensional electron gas.

In the weak frustration regime, the FPS mechanism is not sensitive to
the particular compressibility anomaly and a unified treatment can be
reached. A series a morphological transitions is generally expected
resembling the situation found in other classical systems
\cite{seu95,oht86}. In this limit the phase diagram can be safely
determined by means of a simple uniform density approximation\cite{lor01I,lor01II,lor02,ort06,ort07}.

On the contrary, at strong frustration, two different universality
classes arise.   
In systems with a negative electronic compressibility
region ($\gamma=2$) a critical value of the frustration exist
$Q_{c}$ for both  $D=2,3$. Close to $Q_{c}$ soft inhomogeneities 
appear. They continuously evolve into the sharply defined structures of the weak frustration regime.   

For systems with a cusp singularity in the electronic
compressibility ($\gamma=1$) the system dimensionality
plays a key role. Indeed a maximum frustration exists only in three-dimensional systems. 

According to the maximum size rule, domains cannot have all linear dimensions  much larger than the
screening length. Therefore mesoscopic domains are generally 
expected in systems with small compressibility as bad metals,
systems close to metal-insulator transitions and
systems with very anisotropic electronic properties as indeed found.

\end{section}

\end{document}